\documentclass[a4paper]{article}

\usepackage{INTERSPEECH2020}
\usepackage{amsmath,graphicx}
\usepackage{url}
\usepackage{subcaption}

\title{The phonetic bases of vocal expressed emotion: natural versus acted}
\name{Hira Dhamyal, Shahan A. Memon, Bhiksha Raj, Rita Singh}
\address{Carnegie Mellon University}
\email{hyd@cs.cmu.edu, samemon@cs.cmu.edu, bhiksha@cs.cmu.edu, rsingh@cs.cmu.edu}

\begin{document}

\maketitle
\begin{abstract}
\emph{Can vocal emotions be emulated?} This question has been a recurrent concern of the speech community, and has also been vigorously investigated. It has been fueled further by its link to the issue of validity of acted emotion databases. Much of the speech and vocal emotion research has relied on acted emotion databases as valid \emph{proxies} for studying natural emotions. To create models that generalize to natural settings, it is crucial to work with \emph{valid} prototypes -- ones that can be assumed to reliably represent natural emotions. More concretely, it is important to study emulated emotions against natural emotions in terms of their physiological, and psychological concomitants. In this paper, we present an on-scale systematic study of the differences between natural and acted vocal emotions. We use a self-attention based emotion classification model to understand the phonetic bases of emotions by discovering the most \emph{`attended'} phonemes for each class of emotions. We then compare these attended-phonemes in their importance and distribution across acted and natural classes. Our tests show significant differences in the manner and choice of phonemes in acted and natural speech, concluding moderate to low validity and value in using acted speech databases for emotion classification tasks. 
\end{abstract}

\section{Introduction}
\label{sec:intro}
\emph{Can vocal emotions be emulated?}  This question has led to long standing debates in the speech community regarding \emph{natural} versus \emph{acted} emotions, in the context of emotion classification and emotion categorization tasks. 
To conduct any speech based emotion research, an important factor is the nature of the speech samples or the vocal stimuli, and whether those samples are representative of natural emotions. 
Natural emotions can best be defined as emotions that are spontaneous and involuntary. Acted emotions, on the other hand, are prompted and voluntary. 
Because acted emotions are volitional, researchers argue that the physiological and psychological responses that natural emotions induce are absent from acted emotions
 \cite{jurgens2015effect,wilting2006real}. Nevertheless, research on emotion perception uses acted emotions as convenient proxies for natural emotions. While many past studies have focused on presenting perception tests for natural versus acted emotions with mixed conclusions \cite{wilting2006real,scherer2013vocal,audibert2008we}, there is a lack of an at-scale systematic framework to study the differences and similarities in those classes. 

\begin{figure}[htb]
  \centering
   \centerline{\includegraphics[width=6cm]{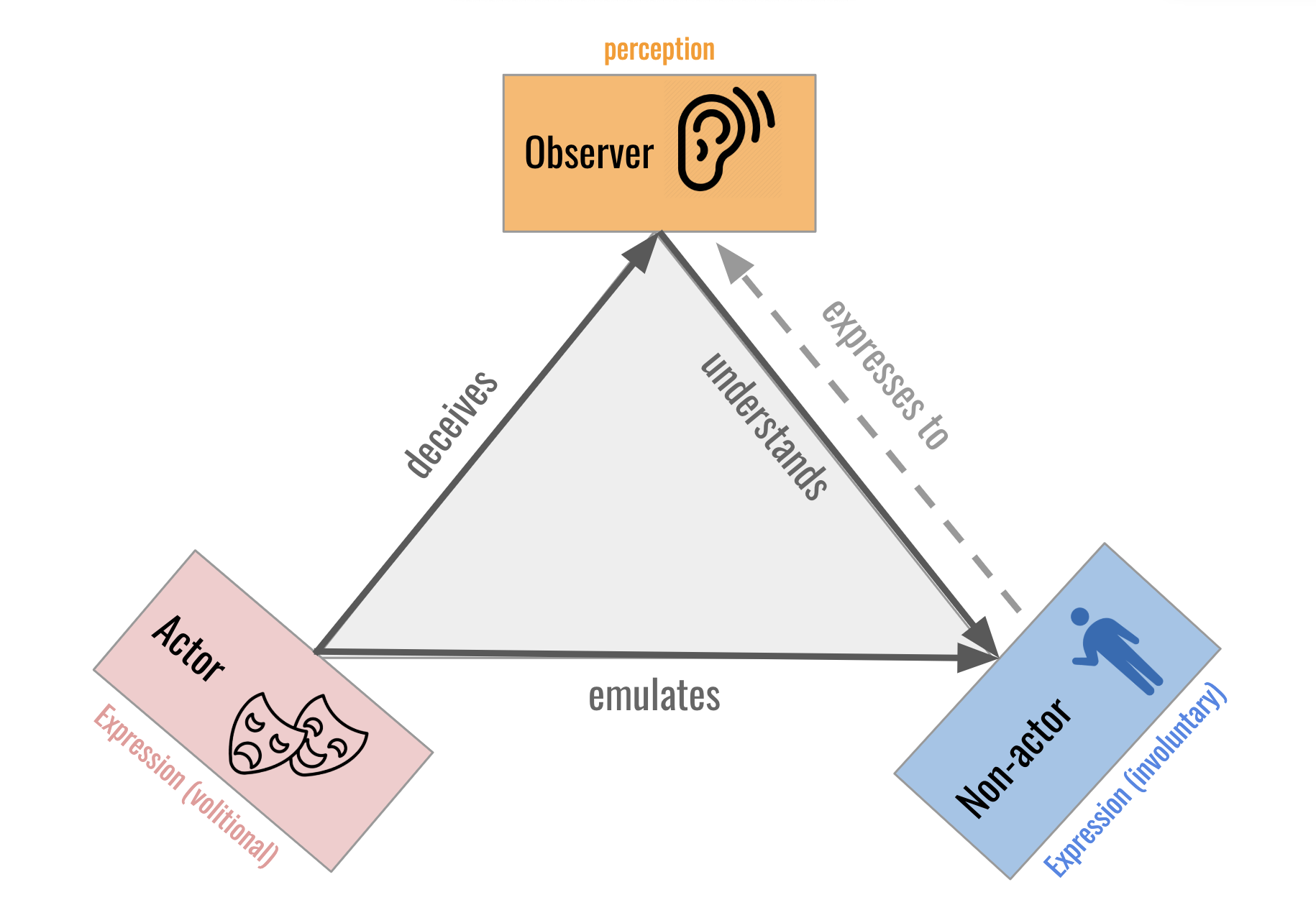}}
  \caption{The non-actor, actor and observer (NAO) framework for the communication of emotions.}
    \label{fig:noa_model}
\end{figure}

In order to develop such a framework, we must recognize that there are, in fact {\em three entities} to be considered.
The communication of vocal emotions is, at its essence, a combination of an encoding and a decoding process. The subject {\em expressing} the emotion encodes their emotional state into the low-dimensional speech signal. The subject {\em perceiving} the signal decodes it to make inferences about the state of the speaker. We will distinguish between two types of encoders:  the {\em non-actor} who actually experiences the emotion, and the {\em actor} who may not. In all cases, the decoder is an {\em observer}, who's only cue in terms of vocal emotions is the vocal stimuli. Based on these, we propose the \emph{non-actor, actor, and observer (NAO)} model (Figure \ref{fig:noa_model}), which represents all three entities and the relation between them. The actor aims to encode synthetic emotion in a manner that the observer cannot distinguish from the genuine emotion encoded by the non-actor. This enables us to formulate a hypothesis that can be formally tested -- that there nevertheless remain identifiable fundamental differences in the encoded signals in the two cases. If the test fails, that would mean natural emotions \emph{can} be emulated, and that acted emotions can be used as proxies for natural emotions. If the test passes, however, that would signal towards dichotomy between acted and natural emotions, leading to a low validity and value in using acted stimuli.


We note that non-actor and the actor differ in their encoding of emotions. Because natural emotions are, for the most part, involuntary, they include physiological and psychological responses as concomitants, such as heart rate, breathing rate, muscle tension, and mood. These physiological changes manifest in the voice by changing the spectro-temporal structure of individual sounds \cite{johnstone1999effects, scherer1977cue}. For example, \cite{williams1972emotions, myers2013inherent} argue that vowels and consonants produced in \emph{fear} are often more precisely articulated than they are in \emph{neutral} situations. The physiological changes along with the psychological factors also define the choice of phonemes (i.e. the lexical content) and prosodic cues \cite{scherer2003vocal, patel2011mapping}. As an example, words of aggressive nature are more likely to be used by an individual in an aggressive mood \cite{whissell1999phonosymbolism}. 
The  encoding of emotion is hence an aggregate, or {\em perceptual sum} of these acoustic, phonetic and linguistic influences, or {\em factors}. 
The observer decodes the perceptual sum of these factors to make their inference about the emotional state of the speaker.


In emulating an emotion, the actor attempts to produce a somewhat similar {\em aggregate} of these factors as the non-actor, {\em i.e.} a combination of factors that he expects the observer to decode into a near identical perceptual sum. If the actor succeeds, he conveys the target emotion to the observer. We hypothesize that in doing so,  the {\em individual factors} that the actor produces will, however, still be incorrect or even implausible, even though the perceptual sum may be plausible. 


The distinction is, perhaps,  best illustrated by the study, ``The President's Speech'', narrated by well-known neurologist Oliver Sacks  \cite{sacks1985president}, in which two types of patients, \emph{aphasic}, and \emph{tonal agnostic}, both find a presidential candidate's posturing on TV to be screamingly implausible, although normal viewers have no problem with it. \emph{Aphasic} patients are highly sensitive to expression and tone, but cannot interpret the words. On the other hand, \emph{tonal agnostic} patients lack any sense of expression and tone, and pay attention to exactness of words and word use to capture the emotion. The skilled actor, in this case the presidential candidate, attempts to convey feelings and emotions through a combination of affect that is given a pass by normal viewers.
However, both types of patients who, unlike normal people, only perceive some of these factors, find them sufficiently implausible as to cause them distress. Simply by their inability to consider the totality of affect that normal people can perceive, the patients cannot be lied to or deceived.



Vocal (or indeed any) expression of emotion is, of course, a complex phenomenon, and the complete set of acoustic, phonetic, linguistic and prosodic factors used in expressing it is still not fully understood. To test our hypothesis, we must nevertheless identify one or more of these factors that can be statistically quantified. As mentioned earlier, physiological and psychological changes concomitant with emotion are known to affect the \emph{choice of phonemes} and their \emph{manner of delivery}. We will refer to these as the \emph{phonetic bases of vocal emotions}. By our hypothesis, there will be a statistically measurable difference in these between the actor and non-actor.

To verify our hypothesis, we require a mechanism to quantiably extract these bases from the speech signal. To do so, we train a neural network model for emotion classification tasks on two datasets, one of natural speech and the other of acted emotional speech, using an \emph{attention} mechanism. The attention aims to identify the most important phonemes in an utterance in order to classify its  emotion. We compare the statistical patterns of the most \emph{attented} phonemes across actor and non-actor. As we will see in the final sections of our paper, these factors do indeed differ in a statistically significant manner, bringing the validity of conclusions drawn from acted emotional speech as a proxy for natural emotion into question.

.

\section{Background Work}

There has been some work done to understand the differences between acted and natural emotional speech. The results found from these studies are somewhat contradictory to each other. Studies like \cite{scheiner2011emotion}, which analyze acted and natural emotional speech with the help of human listeners have concluded that the listeners are not able to distinguish between the two categories. The problem is also studied in the domain of false expression, where the truthfulness of the expressed emotion is studied. It also reaches the conclusion that humans are less likely to differentiate between the two. On the other hand, studies like \cite{audibert2008we}, which also use human listeners, conclude that about $78\%$ of listeners were able to differentiate between the natural and acted emotion with only audio clues and even more could differentiate when provided with audio-visual cues.

The above studies primarily analyze the effect of acted emotion on the listener. In our work we do not consider the listener (observer in the proposed NAO model) to be  a valid discriminator between acted and natural emotion, hypothesizing instead that the factors that comprise natural and acted emotion differ significantly irrespective of the observer's response. The following studies supports our hypothesis although not within a systematic framework to analyze the difference.
\cite{truong2009does} concludes that acted and natural speech innately differs based on voice quality. Acted speech is considered to be delivered in a more emotionally intense fashion but also that acted speech affects the vocal expression in a more general way, without the nuances of the changes caused by the natural emotion \cite{jurgens2011authentic}.
Some studies have focused on only particular aspects of vocal emotion like \cite{audibert2010prosodic} which concludes that the two are different based on the prosodic properties of the speech.


\section{Neural Model}

In order to extract the phonetic bases of vocal emotion, we propose a neural network model. We design the model to take into consideration both the lexical and acoustic aspects of the utterance and also the relationship between the two, to capture the phonetic bases of emotion. The linguistics should guide the model about the important parts of the acoustic. To create a vector representation of the linguistic part of the input, we pass it through an LSTM which captures the contextual information of the linguistics. This forms a context-sensitive lexical vector.

\begin{figure}[htb]
    \centering
    \includegraphics[width=0.25\textwidth, height=0.3\textheight]{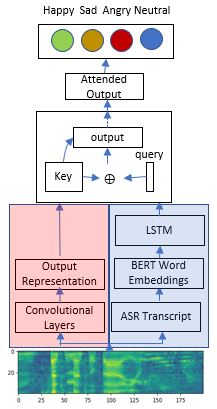}
    \caption{Neural network model used for the emotion classification with attention mechanism}
    \label{fig:model_diagram}
\end{figure}

To capture the relationship between the two modalities, we utilize an attention-based mechanism. This enables the context-sensitive lexical vector to put attention on some parts of the audio, forming importance weights. The weights, when applied back to the input audio, make the output high in parts that the lexical vector points to and others become low in value. 
A feature vector is created from this weighted output, which thereafter goes into the classification layer for emotion. Training the model maximizes the classification accuracy, but in doing so, it teaches the model to create feature vectors which would be differentiable for the emotion classes. This, in turn, optimizes the attention mechanism, thereby allowing the lexical vector to focus only on those parts of the audio which would lead to the highest classification accuracy. This lays the basis of the model we have used, as shown in Figure \ref{fig:model_diagram}. 

Since we need both the acoustic and lexical content of an utterance, to train the model, we require the transcription of the recording. To get the transcription, the recording is passed through ASR. We used Google API \cite{googlespeechtotext} to extract the transcription. Each word in the transcript is represented as a BERT \cite{devlin2018bert} contextualized word embedding. These embeddings are passed through an LSTM layer (shown in blue). 
For the attention layer we represent the keys as the output of the convolutional layer; a 3-dimensional output. The query is the last hidden state of the LSTM passed through a linear projection.

The network is optimized using the Cross-Entropy loss, with weights for individual labels due to the class imbalance in the datasets. 
The ASR based transcript of the utterance is segmented into different phonemes using an HMM based phoneme segmentor \cite{lamere2003cmu}. 
Once the model is trained, the attended-phoneme outputs are inspected.

\section{Experiment}


\subsection{Dataset}
We use two types of datasets; acted and natural. 
We run our experiments for only four emotions: angry, happy, sad and neutral. 
\subsubsection{Acted Data}
The acted dataset used in the experiment is IEMOCAP \cite{busso2008iemocap}. It consists of ten sessions, each of which is a conversation between two actors. The conversations are divided into labeled sentences. 
We implement a 10-fold cross validation training setup. In each fold, data from 9 speakers is used for training the model and data from 1 speaker is used for testing. 
The data consists of $1103$ angry, $1636$ happy, $1708$ neutral and $1084$ sad utterances. The average duration of utterances in this dataset is $4$ seconds. 
\subsubsection{Natural Data}
For  natural-speech, we used the CMU-SER data \cite{perceptionsei}. This dataset has been collected from NPR podcasts \cite{npr}, and television programs hosted by the Internet Archive \cite{internetarchive}. The dataset is annotated using the Amazon Mechanical Turk \cite{turk2012amazon}. It has $6000$ utterances in the training set and  $2571$ utterances in the test set, with a total of $1099$ angry, $3028$ happy, $1262$ neutral, and $611$ sad utterances. The average duration of utterances in this dataset is $5$ seconds.
Further details of the CMU-SER dataset can be found in \cite{perceptionsei}. 

\subsubsection{Alleviating speaker-dependent bias}
Because we compare acted versus natural emotions based on the two datasets with difference in speakers, it is possible for our phonemic content and, hence, phoneme distributions and the attended phonemes to be influenced by the word choices of different speakers. To ensure that our analyses only reflect the differences in the emotional content rather than the differences in speakers, we eliminate speaker dependencies at the time of training our model. Because the natural dataset is collected from a diverse set of online sources, it is reasonable to assume that there are fewer cases of a speaker represented more than once in the data. On the other hand, the acted dataset consists of 10 speakers only. Hence, we perform leave-one-out cross validation to alleviate speaker dependencies in the results. These steps ensure that our models and analyses are robust to the difference of speakers.
\subsubsection{Characterizing content-dependent bias}
\begin{figure}[htb]
  \centering
  \includegraphics[width=1\linewidth,height=0.45\linewidth]{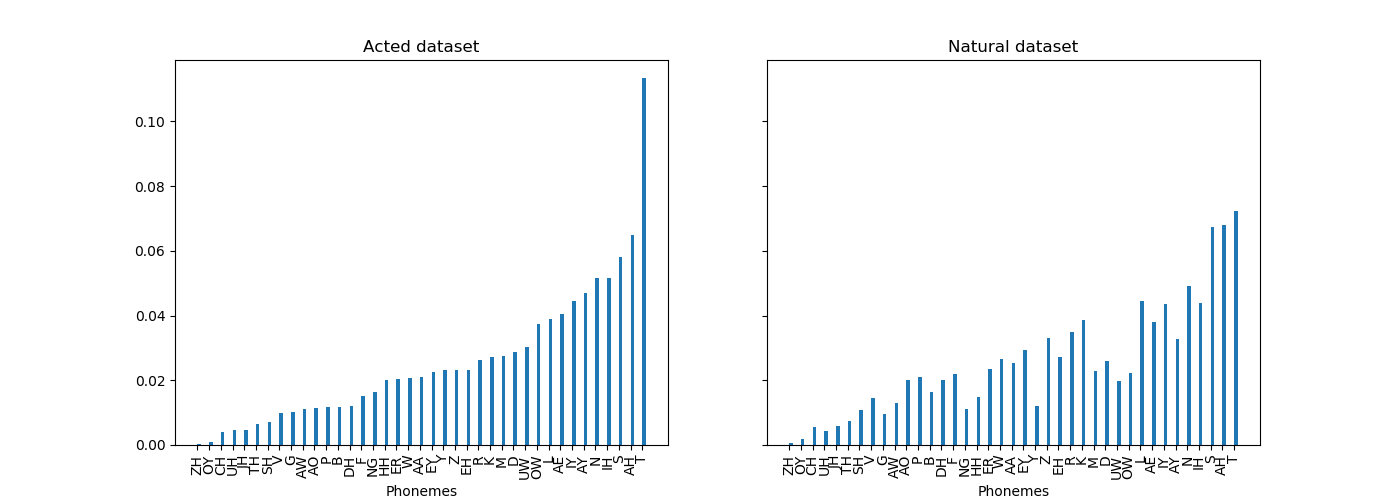}
  \caption{Total phoneme distributions under both natural and acted dataset}
    \label{fig:total_phone_distr}
\end{figure}

It is also possible for our analyses to be influenced by the differences in the content of the two datasets. To ensure that the difference in content is not a confounding factor, we study the phoneme distributions of the two datasets. Figure \ref{fig:total_phone_distr} presents the phoneme distributions of both datasets. To determine if the difference between the distributions is statistically significant, we run a \textit{Wilcoxon rank test} \cite{wilcoxon1992individual}. The Wilcoxon rank test is a non-parametric test, used to compare two related samples. 
In this case, the null hypothesis $\mathcal{H}_0$ is that there is no difference in the distributions of the phonemes under the two datasets, and the alternative hypothesis $\mathcal{H}_1$ is that there \emph{is} a difference between the distributions of the two datasets. We obtain a p-value of $.3$, therefore with $\alpha=.05$, we fail to reject the $\mathcal{H}_0$. This ensures that the difference in the phonemic content of the two datasets is unlikely to affect the distribution of the attended-phonemes. 
\section{Analysis and Results}

We base our results on the output of the attention mechanism from the neural model described earlier. Since the model is trained for emotion classification over four emotions, it is useful to note the classification accuracies achieved by the model over the two datasets. On the acted data, we achieve a classification accuracy of $72\%$ and on the natural data, we achieve a classification accuracy of $52.4\%$. 

\begin{figure}[]
  \centering
  \centerline{\includegraphics[width=0.9\linewidth,height=0.7\linewidth]{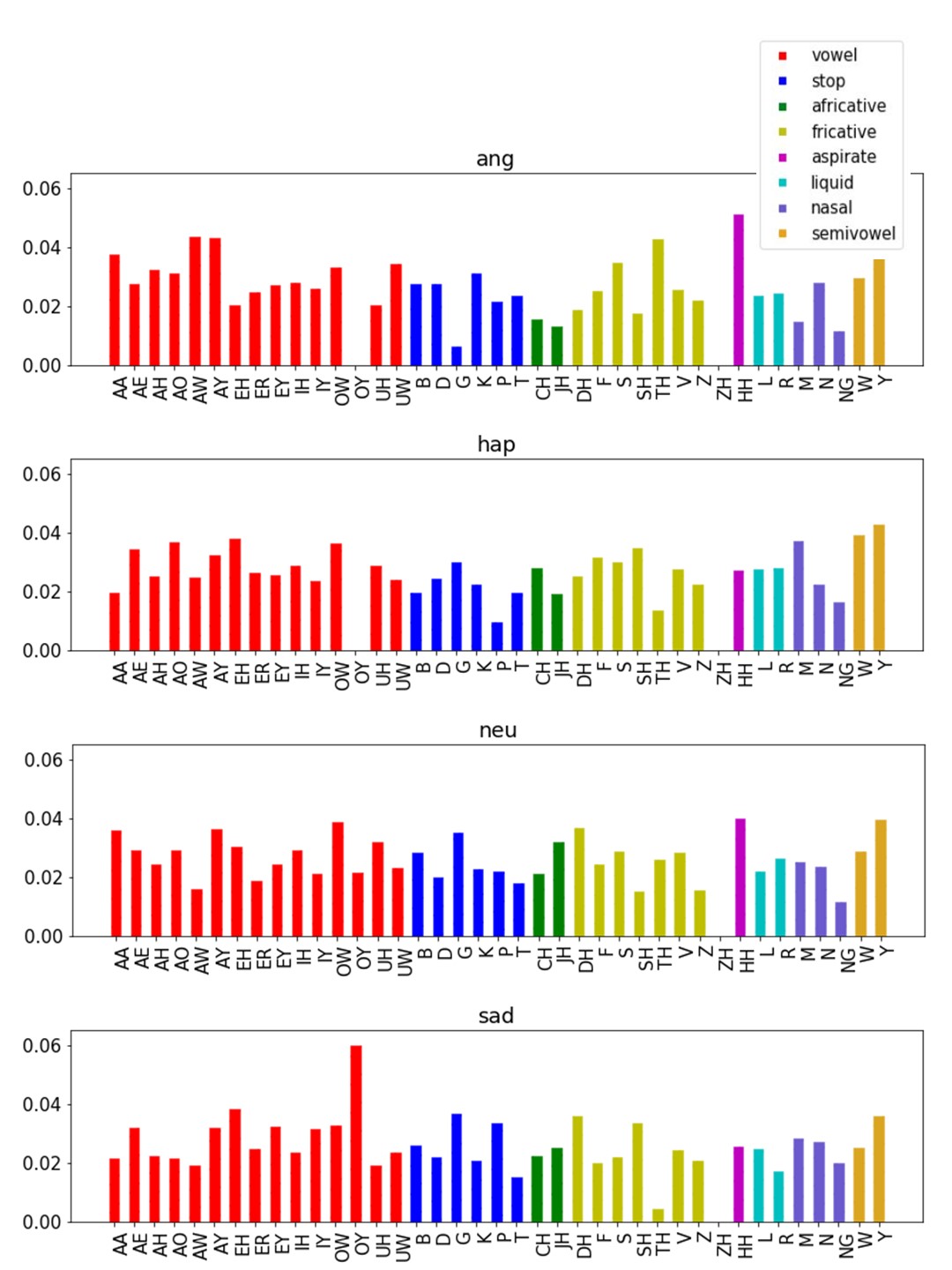}}
  \caption{Distribution of attended-phonemes (and phonetic groupings) across four emotion classes on acted data}
    \label{fig:phone_dist_acted}
\end{figure}

\begin{figure}[]
  \centering
  \centerline{\includegraphics[width=0.9\linewidth,height=0.7\linewidth]{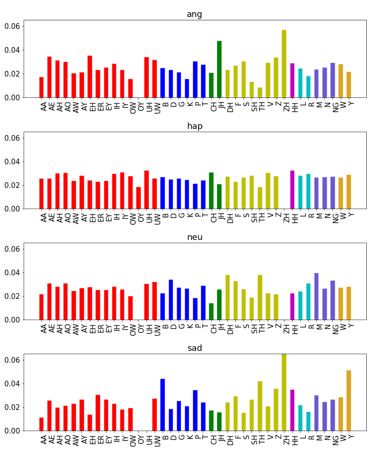}}
  \caption{Distribution of attended-phonemes (and phonetic groupings) across four emotion classes on natural data}
    \label{fig:phone_dist_natural}
\end{figure}

To perform analysis on the attended-phonemes, for each emotion we aggregate the phonemes with the highest attention output. We normalize their frequencies by the total frequency of the phoneme in the data. Figures \ref{fig:phone_dist_acted} and  \ref{fig:phone_dist_natural} show the distributions of these attended-phonemes for acted and natural conditions (from the corresponding datasets) respectively, for each emotion. We note several differences between the two distributions. The frequency of fricatives and stops is higher in natural speech than in acted speech.  
We also observe that the frequency of vowels is higher in acted speech than in natural speech. Specifically, the phonemes /AA/, /B/ and /IH/ occur more frequently in acted speech. Moreover, an overall higher percentage of nasal phonemes occurs in natural speech.

\begin{figure}[!htb]
  \centering
  \includegraphics[width=0.8\linewidth,height=0.3\linewidth]{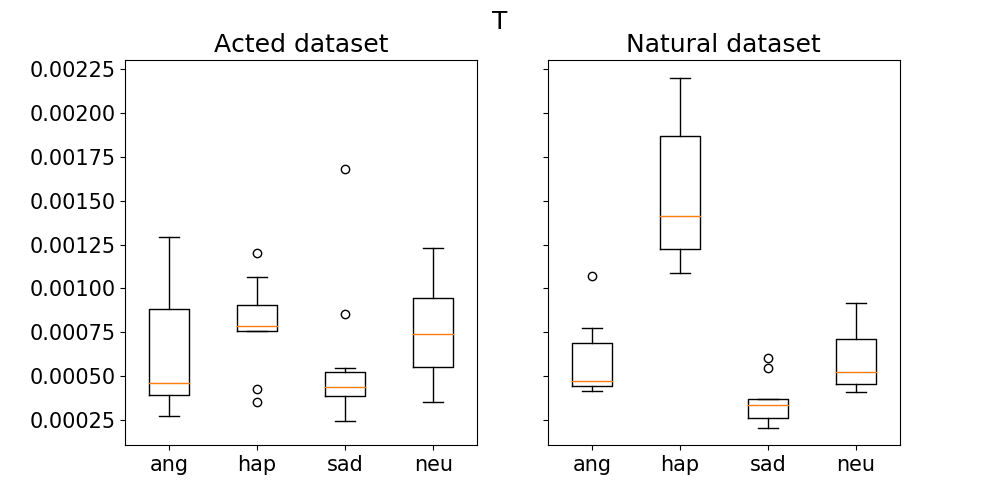}
  \includegraphics[width=0.8\linewidth,height=0.3\linewidth]{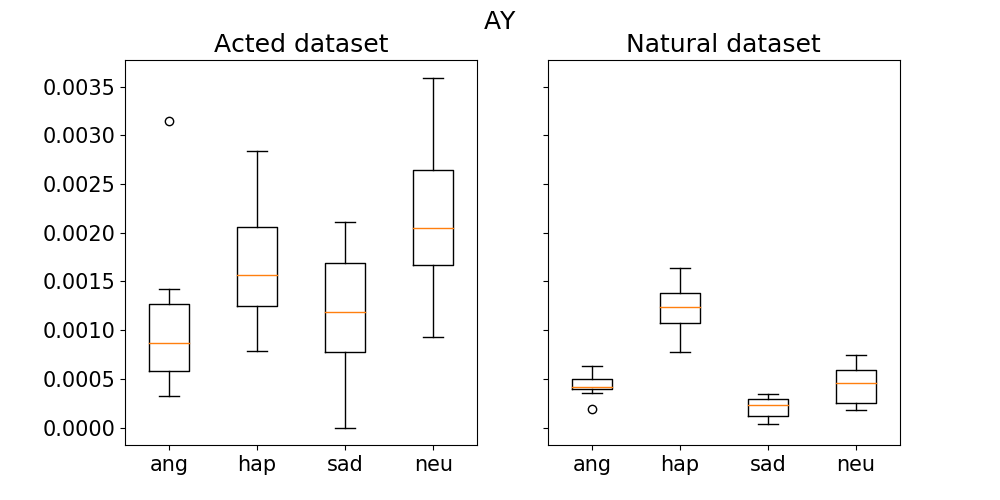}
    \includegraphics[width=0.8\linewidth,height=0.3\linewidth]{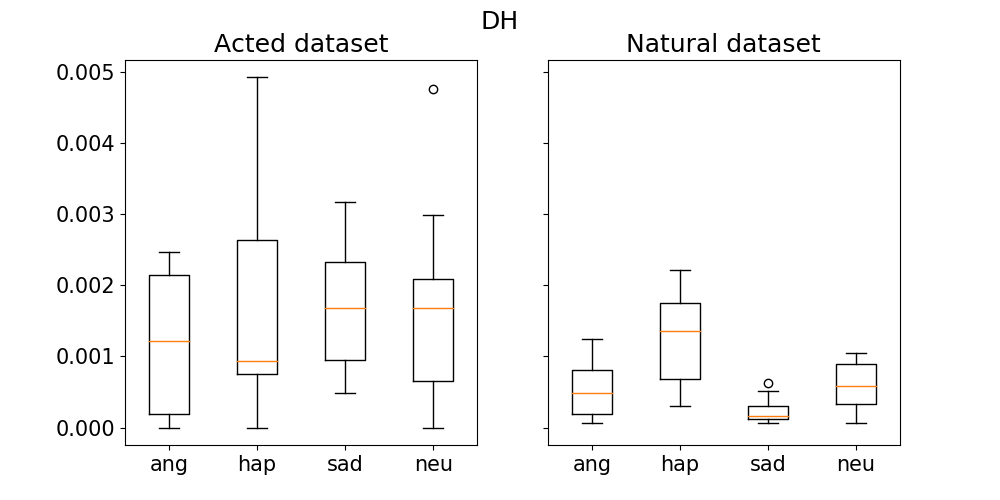}
  \includegraphics[width=0.8\linewidth,height=0.3\linewidth]{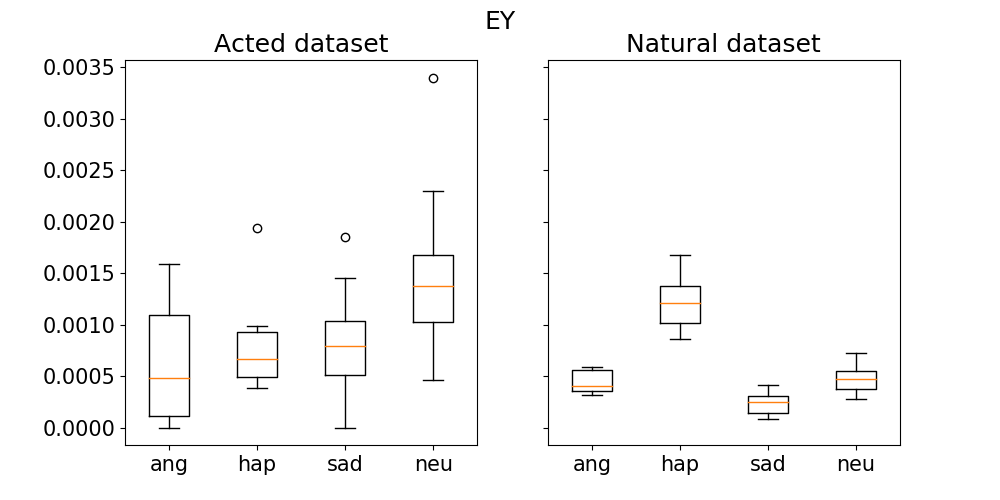}
  \caption{Box plot of attented phoneme /T/, /AY/, /DH/, /EY/ in natural versus acted dataset. The figure highlights the frequency difference among the two datasets. Note the median values for the box plots are very different for both datasets for a given emotion.}
    \label{fig:phone_IY_AY}
\end{figure}

We also study the attended-phoneme distribution under different test subsets created for the 10-fold cross validation procedure to ensure consistency of the attended-phoneme distribution within the dataset. Variation of the phoneme frequency from the 10 different cross validation results are shown in the box plots in figure \ref{fig:phone_IY_AY} for both datasets. It can be observed that the results have lower variation in natural speech than in the acted speech. In general, the same variation trends hold for other phonemes as well. 


 

The box plots also illustrate the difference in the frequencies of each phoneme in the two datasets. In particular, we observe the frequencies of the vowels like /IY/, nasal phonemes /M/, /N/, stop phonemes /T/, and fricatives like /DH/ (figure \ref{fig:phone_IY_AY}) to be different among the two datasets. 
To calculate the significance of these differences for all emotions, we run a standard t-test. We find a \textbf{statistically significant result for all four emotions $(p<.05)$}. 
Therefore, in the context of natural versus acted classes, our analysis concludes that there are significant differences between the phonetic bases of the two classes. Consequently, we  conclude moderate to low validity and value in using acted emotions as proxies for natural emotions, suggesting that  researchers should be wary of arriving at conclusions about natural emotions using acted emotion datasets. 

We would like to note that this study has only inspected English language data. However, the framework provided can easily be applied to any other language. We leave the investigation of the phonetic correlates of emotion in other languages, and its comparison with the conclusions provided in this study, as a possible future work. 

One limitation of this study is the lack of the same set of observers across the acted and natural datasets. While we have no control over this within our analysis, given the diversity of observers for the two datasets, we expect little statistical observer bias. However, this remains to be verified by future studies.




\section{Conclusion}
In this paper, we present a study of the differences observed between natural and acted emotion with respect to their phonetic bases. Phonetic bases of emotion comprise `what' phonemes are used and the `manner' they are delivered in to express the emotion. 
To run a quantifiable test, we model the task as an attention-based emotion classification problem. The attention mechanism aims to capture the ``attended'' phonemes in order to get the correct classification. We then calculate the distribution of these attended-phonemes, and examine how their distribution varies between natural versus acted emotions. We observe several differences, for example, a higher occurrence of fricatives and stops in natural speech than in acted speech. We obtain statistically significant difference in the attended-phoneme distribution among natural and acted emotion. Therefore, our hypothesis stands true. The differences in phonetic bases signal towards a dichotomy between natural and acted emotions. This study has applications in speech emotion recognition, emotional speech synthesis, and human computer interaction. The approach taken in this paper, i.e, exploiting the dynamics of the neural model, allow us to not only use it for marking distinctions between acted versus natural speech, but also to apply it to other problems, such as of exploring the phonetic bases of voice disorders, e.g vocal palsy. 

 \pagebreak 
\bibliographystyle{IEEEtran}

\bibliography{mybib}

\end{document}